# The bioacoustic proof of the effects of raising awareness of noise pollution among visitors using binding communication

David REYMOND*, Daphné DUVERNAY[1], Frédéric ELY[1] and Hervé GLOTIN[2]

[1] *University of Toulon and Aix-Marseille University, EA 7492 IMSIC (Institut Méditerranéen des Sciences de l'Information et de la Communication), Université de Toulon, Toulon, France.*

[2] *University of Toulon and Aix-Marseille University, LIS – Laboratoire d'Informatique et Systèmes – UMR 7020, Université de Toulon, Toulon, France.*

*Corresponding author: *david.reymond@univ-tln.fr*

*Résumé*. **Sensibilisation des visiteurs du Parc National de Port-Cros aux nuisances sonores par la communication engageante et preuve bioacoustique des effets**. En supposant que l'impact anthropique des nuisances sonores occasionnées par les visiteurs d'un parc national pouvait être réduit par des actions de communication, nous avons élaboré en collaboration interdisciplinaire associant bioacoustique et communication, un protocole de mise en évidence de l'impact anthropique associé à un protocole de mesure d'efficacité des actions de communication. Une communication d'éco-sensibilisation simple sera comparée à une communication de sensibilisation engageante. Nous traitons les questions suivantes : est-ce que la sensibilisation des visiteurs les conduit à modifier leurs attitudes et comportements en vue de limiter leurs propres nuisances sonores ? Est-ce que l'engagement améliore la sensibilisation ?

Nous détaillons ici les principaux concepts convoqués et la méthodologie suivie afin de présenter les premiers résultats et les travaux en extension.

**Mots clés : protocole interdisciplinaire, communication engageante, bioacoustique**

**Abstract.** Assuming that the anthropogenic impact of visitors to a National park can be reduced by communication actions, we have drawn up a bioacoustical protocol combined with a protocol to measure the effectiveness of the communication actions. A simple eco-awareness communication method will be compared to a binding communication method. We address the questions: does visitors' eco-awareness drive them to change their attitudes and behavior in order to limit their own noise disturbance, and does commitment improve their awareness? We detail here the main concepts considered and the methodology followed to discuss the preliminary results.

**Introduction**

The Port-Cros National Park is located in the south of France, Provence, North-Western Mediterranean Sea and was founded in 1963 (Barcelo et Boudouresque, 2012). The National Park is endowed with a Scientific Council since 1964 with a specific role: to develop knowledge of the territory as a target of protection and to provide answers to questions from the managing authorities (C. F. Boudouresque *et al.*, 2013). At the beginning of 2016, Port-Cros National Park (PCNP) launched a debate and calls for projects to assess the tourism carrying capacity (TCC[1]) of the islands of Porquerolles and Port-Cros (Deldrève and Michel, 2019). With the aid of these studies, the Park is seeking to establish different scenarios for the sustainable economic development of its territory, based on hypotheses regarding the evolution of tourist numbers (Brécard and De Luigi, 2016; Van Der Yeught, 2018). The characteristics of the site, due to its particular biodiversity (Médail *et al.*, 2013), make it a showpiece for the national strategy for the conservation and sustainable management of the Mediterranean coast. However, space is limited and the biosphere can only thrive in harmony with local and tourist communities. This biosphere appears threatened by the poorly controlled development of human activities, the critical level of biological invasions, overfishing (C.-F. Boudouresque et al., 2017; Maxwell et al., 2016) or marine pollution but, we guess, the excessive number of visitors to the site at certain times, jeopardising its ecological balance.

This situation has led the PCNP to seek solutions with the support, among others, of the scientific community of the University of Toulon, through cross-disciplinary scientific platforms, bringing together all the university's laboratories.

Our research team, made up of five research-lecturers, from information sciences (IT and bioacoustics) and from Information and Communication Sciences (ICS), joined forces with a collective project[2] created to target two main objectives: determinate a way to identify ecosystem's activity disturbance using soundscape and, secondly, try to provide a way to proof sensibilisation and communication actions efficiency. We address mainly the second objective in this article.

This paper comprises three parts: a summary of the main concepts included in our research, the experimental methodology, and finally, a discussion of the preliminary results.

---

[1] The concept of TCC, originating in the scientific fields of demographics and biology, is used in research as a tool for creating sustainable tourist development strategies. All geographic areas offer limited resources and withstand activity below a threshold which must not be exceeded as the overexploitation of the resources of an eco-system always results in the destruction of part of those whose habitat it is. This threshold is commonly called the carrying capacity and applied to a tourist destination it becomes the TCC

[2] The Captîles project involves 5 laboratories and 23 researchers in separate fields (ICS, management, economics, biology, IT and oceanology).

# The central question of the measurement of ecological awareness

## The PCNP: a field of experimentation for ICS research

One of the mission of National Parks is to ensure the conservation of natural eco-systems, without overlooking their tourist vocation, but at the same time respecting the rules inherent to sustainable development and ecology (Larrère and Louisy, 2013; Sumaila *et al.*, 2000). These national parks thereby offer opportunities for research and experimentation with regard to raising the general public's awareness of environmental conservation.

In order to secure conditions favourable to sustainable tourism in these natural spaces, it is essential to encourage visitors to commit to behaving responsibly in terms of the protection of the environment.

## An innovative combination of bioacoustics and binding communication

During the Second World War, a period of food shortages, Lewin (1947) showed the limitations of information and awareness campaigns on changing eating habits. While a series of conferences aimed at raising American housewives' awareness of the war effort had helped to improve their nutritional knowledge, it had done nothing to change their cooking habits, offal being considered particularly unpalatable in American society. Lewin then had the idea of getting housewives to publicly commit to cooking these inexpensive cuts of meat once back home, by asking them to express their willingness through a show of hands in front of the group after they had discussed amongst themselves the nutritional benefits of these foods. Following this simple public gesture, 32% of them adopted the desired behaviour by changing their eating habits; this was 10 times more than with the basic awareness campaign. Thereby, between the idea and the behaviour, Lewin (1947) showed the determining factor of the decision which would bind the individual to his or her actions, what the researcher would call the 'freezing effect'.

Later, similar observations were made in the field of health (AIDS and smoking prevention): health information campaigns alone are not enough to encourage students to use a condom (Morin, 1994) or to persuade young people not to smoke (Peterson *et al.*, 2000). In other words, convictions do not always result in good practices, or consequently, attitude does not systematically guide behaviour.

Joule and Beauvois (2005) discussed the path from persuasive communication to binding communication. The authors review methods encompassed in the 'free will compliance paradigm' (low-ball, teasing, foot-in-the door, touch, and 'you are free to'), reaffirming that they increase the probability of convincing others to comply to one's requests by allowing them to feel free to do so. Hence, they introduce the theory of commitment from a social psychology perspective in order to present 'binding communication' as an expression of a persuasive communication theory comprising both commitment and free will compliance.

In information and communication sciences, Bernard et Joule (2004) then Bernard (2006) integrated into their theory of binding communication the conceptual and methodological principles of these paradigms, themselves originating in the 'commitment theory' (Kiesler et Jones, 1971) in social psychology, which shows that individuals are more likely to commit on the basis of their actions than of their values. Joule and Beauvois (2002, 2005), subscribing to this assumption, thereby emphasised that it is not the individual committing on the basis of his or her personological characteristics, but the situation which is binding the subject on the basis of his or her objective characteristics (Joule et Beauvois, 2002, 2005). Binding communication thereby constitutes an appropriate theoretical and methodological tool in the context of the commitment of users of the PCNP to behave in such a way as to respect the island's wildlife. It is also an innovative tool in the sense that it enables our bi-disciplinary team of researchers to put in place complementary acoustic data recording experimentations. Our research theme asks if raising visitors' awareness lead them to change their attitudes and behaviour in order to limit the noise pollution they cause and if the use of binding communication improve this awareness?

Our work firstly focused on the observation of the noise (voices, footsteps, radio, etc.) footprint of visitors, a source of disturbance for wildlife, an element which has not been sufficiently taken into account by stakeholders. Whether it be as we walk around or relax on the island, the coming and going of pleasure craft and transport boats, etc., visitors generate noise which would pollute the life of marine and land animals. But to what extent?

Secondly, we created a communication tool to encourage visitors to commit to adopting quieter behaviour as they move around the site.

## Methodology and experimentation of two interdependent research protocols: bioacoustics and binding communication

Over three consecutive days in September 2016, we carried out on the PCNP site itself, an experiment in three distinct phases (Fig. 1). The noise pollution caused by visitors to the Park was measured using a network of microphones; before and after the introduction of a binding or non-binding eco-awareness communication protocol. This experimental approach thereby enabled us to test the hypothesis that attitudes and noise behaviour can be changed through the adoption of the binding communication protocol combining awareness and commitment methods.

Our field methodology comprises three distinct phases (Fig. 1):

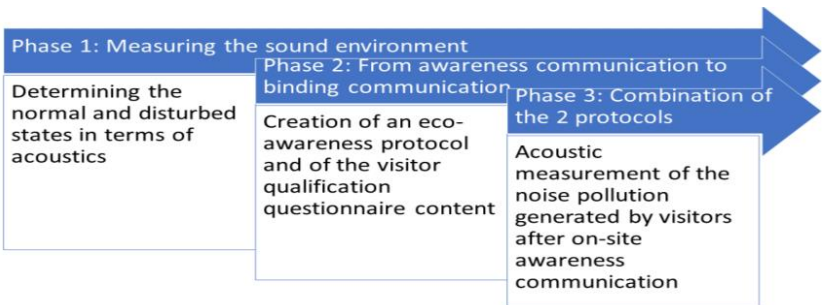

***Figure 1:*** *Research protocol in three steps*

Each of these three phases is described below: the first phase concerns primarily the IT/bioacoustics IT & Systems laboratory, the second phase the ICS laboratory, while the third and final phase involves both laboratories and is carried out on one of the island's three pre-selected 'experimental trails' (Fig. 2). A double experimental protocol is then set up and applied to 3 experimental groups (1 control group, 1 awareness/information group and 1 individual commitment group).

**Phase 1: Monitoring of the acoustic environment**

For the bioacoustics researchers, the approach comprises two stages:

- Acoustic recordings. These aim to evaluate in decibels and in frequency the levels of the bioacoustics environment at the heart of the PCNP, i.e. all the wildlife and plants which make up its acoustic character. The PCNP soundscape is the "result of the permanent intertwining, over the course of time, of two unique histories: natural history and human history nestled between land and sea" (Gérardin, 2013). These recordings comply with the control of major physical factors regulating phonation.

- Correlation of the above values with the number of visitors and their awareness, and study of the quantification of a threshold of visitors above which disturbance of the wildlife is too great, i.e. its levels of bioacoustic emissions are abnormal comparatively when nobody is present.

In order to control the sampling bias at the time of gathering and recording the data, the conditions must guarantee what we called the 'impermeability' of each experimental trail, meaning that they are used exclusively by visitors having been made aware beforehand of the effects on wildlife of the noise they generate.

These conditions require that:

- The flow of visitors be controlled qualitatively and quantitatively on each observed trail;

- The visitors taking the experimental trail only travel in one direction, without being able to walk back the way they came, take any side paths, etc.

- 100% of the visitors to an experimental trail segment (we set a minimum of 100 per session of recording the observed population) have been made aware beforehand of the issue of noise pollution (with the exception of the control group to whom no eco-awareness information had been given for comparative purposes, as the experimental method stipulates)

Our network of sensors was deployed in strategic areas of the island (trails leading to the beach) (Fig. 2) allowing us to gather acoustic data. These sensors enabled us to identify locally the acoustic environment, representative of the park's biodiversity, as well as the noise pollution generated by visitors.

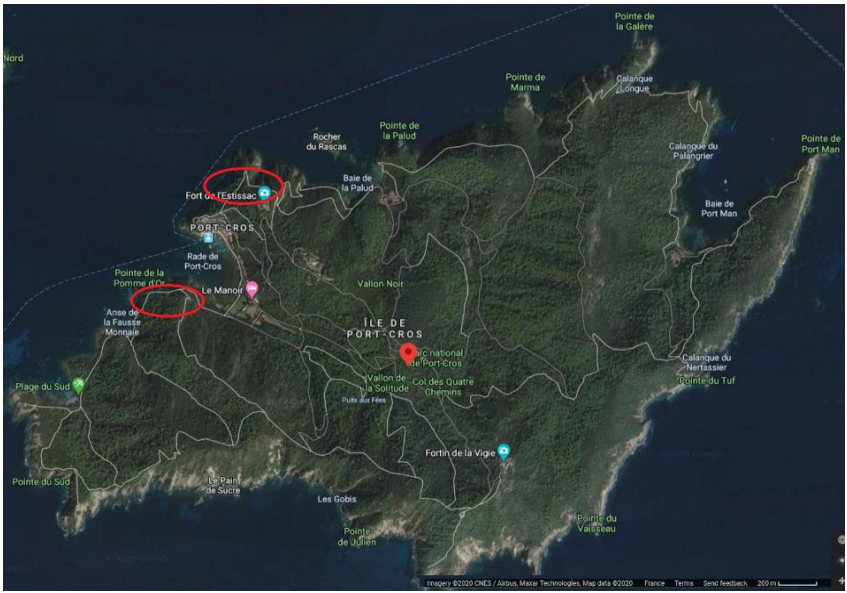

***Figure 2:*** *Satellite image of the island of Port-Cros with the location of the experimental trails (Source: Google Maps). The upper trail used in 2016 and again in 2018 along with the lower trail. Visitors mainly go from the port (between the two locations) to the beaches using these path (« baie de la Palud » or « plage du sud »). See figures 5 and 6 for details on paths and sensor positioning.*

In parallel with this technical tool, we were able to put in place our eco-awareness protocol, in order to test its effectiveness *in situ*.

**Phase 2: From awareness communication to binding communication**

In this second phase, visitors were subjected to three different experimental conditions:

- D1: A group was given 'basic awareness' information without being asked to make a commitment. This 'non-committed' group comprised 52 men and 47 women (4 did not specify their gender) and received a simple 'awareness' questionnaire, comprising questions relating to the impact visitors can have on a national park such as that of Port-Cros.

- D2: Another group was given 'binding awareness' information. This group of 'committed' visitors comprised 47 men and 52 women (4 did not specify their gender in this group either) and received another version of the questionnaire, called 'binding', designed to induce their commitment: it comprises the same questions as the questionnaire given to the 'non-committed' group, but the visitors are this time asked to agree to make less noise by ticking a specific box at the end of the questionnaire.

- D3: A final 'control' group also of around one hundred visitors, visited the trail without being given any awareness information, either basic or binding.

Our protocol thereby consisted in welcoming visitors to the most appropriate of the three predefined trails (A3/B3, Fig. 2), i.e. the one which best meets the bioacoustics calibration conditions (cf. phase 1 above), and inviting these same visitors, in various ways, to limit their noise pollution. To do this, the questionnaire we gave them also contained explanations about the need to make as little noise as possible (by moving slowly, nor shouting, etc.) in order to limit the potential impact it can have on the wildlife. This questionnaire also enabled us to gather the social characteristics (age, gender, level of education, etc.) of the visitors.

The questionnaire therefore consisted, in the context of our experimental approach, in a preparatory action to the commitment to behave in such a way as to respect the soundscape of the PCNP.

**The questionnaire as a preparatory action**

To raise visitors' awareness and gain their commitment and given the logistical context (a forest trail), we decided to use a questionnaire on the environmental impact as a preparatory action. To maximise the effectiveness of this preparatory act, Bernard (2006: 74) , drawing on the work of Burger (1999), lists six characteristics of the context to put in place. We will discuss them one by one, explaining how we adapted them to our study:

1. "The preparatory behaviour must be effectively carried out (it is not enough to state an intention of behaving in a certain way)." We designed a questionnaire of five questions (not including questions about signs and the commitment box to tick for the 'binding' version) to find out:

- How often users of the PCNP taking part in this study visit the park
- Their motivation for respecting the park's regulations
- Their knowledge of the consequences of visits on the plants and wildlife
- The perception of their impact in terms of noise pollution

2. "It is good to help the person establish a link between what he or she has just done and what he or she is by using labelling (internal attribution)." This notion of labelling is important as according to Grandjean and Guéguen (2011) labelling consists in "giving the individual who has just displayed a certain behaviour, information enabling him or her to attribute to himself or herself a positive personality trait related to this behaviour." This labelling was achieved through the binding version of the questionnaire, by formulating the commitment request in the first person: "By ticking this box I am making a personal commitment to be as quiet as possible as I move along the trails of the Port-Cros National Park today." By agreeing to tick the box, the person is labelling themselves as a visitor who respects the environment and is willing to listen.

3. "There must be a certain cost to the preparatory behaviour."

The act represents a relatively high cost and investment in terms of time and energy: filling in the questionnaire takes around 7 minutes (taking into account the need to rank 12 items on certain questions, and the fact that it is not the most comfortable setting for filling in a questionnaire), which is by no means insignificant as visitors are generally in a hurry to get to the famous La Palud beach.

4. "The preparatory behaviour and the behaviour which constitutes the final request must come under the same identification of the action."

The process of action identification described by Wegner and Vallacher (1986) and used by Bernard *et al.* (2007) in the binding communication theory, shows that we can be committed up to a given level of identification. For Wegner and Vallacher (1986), the action is a concept which refers to "the events to which people make reference when they say they are doing something". They thereby distinguish two different levels of linguistic statements which correspond respectively to the 'how' and to the 'why' the action is being carried out. Describing or naming an action by (Wegner and Vallacher, 1986) simply explaining what we are physically doing (e.g.: "I am switching the light off by pressing on a button") corresponds to a 'low level' statement or identification, while explaining the reasons for our action by linking it to the project we are pursuing (e.g.: "By switching the light off I am saving energy and I am helping to protect the environment") corresponds to a 'high level' statement or identification. This cognitive process at work in the language statement which serves to identify the level (low or high) of an action is also a determining factor of commitment. A higher-level identification action is thereby more favourable than a lower level action to the maintaining or perseveration of the commitment to behave in a given way.

In the context of our own experiment, the preparatory act to the commitment with the filling in of the questionnaire (Reducing the noise disturbance of wildlife), enables the action ("I am not making any noise") to be identified at a high level ("I am helping to protect the PCNP, I am helping to eliminate environmentally destructive behaviour").

5. "It is preferable that it is not the same person who asks for the preparatory behaviour and who formulates the final request."

On this point, we refer to the 'methodological pluralism' principle advocated by Bernard and Joule (2004) as part of their ICS research into binding communication. This position aims to "promote, in the field of ICS, the perspective of science in action in society.

We therefore endeavoured to demonstrate, to visitors to the PCNP, our own commitment to this type of approach involving the cooperation of heterogenous players in a project combining communication and action. When handing out the questionnaires, wearing t-shirts with our university logo, we thereby displayed to the visitors how our bi-disciplinary team of university researchers was contributing to the site conservation project commissioned by the PCNP. While the Park remained the project sponsor, the research undertaken at its request should result in the design of "*communication interventions*" of significant "social utility" (in the words of Bernard and Joule) associated with a rigorous scientific approach.

6. "The preparatory behaviour should not be linked to any financial compensation, and more generally, to any promise of reward."

The theory of commitment stipulates that rewards of any nature whatsoever, and likewise any threats of punishment, result in the withdrawal of the individuals. The context of liberty is respected, the visitor is free to choose whether to take part or not.

### Phase 3: Combination of both protocols (phases 1 and 2)

The research protocol is bi-disciplinary. We collected the data on 12 and 13 September 2016: positioned near the A3/B3 trail (Fig. 2), we presented over the course of these two consecutive days both questionnaires to more than one hundred people in each of the two experimental conditions. A total of 206 questionnaires were thereby collected. On the first day we used the basic eco-awareness questionnaire, while the second day was devoted to 'binding' eco-awareness. The analysis of the control group (where the sensors recorded the acoustic data without the presence of the researchers) had been planned for 14 September 2016, but, as bad weather was forecast for that day, it was brought forward to 13 September. Indeed, on this day, after having collected the one hundred or so binding eco-awareness questionnaires (between 8:00 and 11:00), we made the most of our remaining time (11:00-13:00) to set up the observation of the control group. In the end, the data collected for this group could not be used for comparison with those of the experimental groups, as the conditions of the field measurements for this group were consequently not optimal as less people came during this period as well as soundscape differs from earlier time in the same zone.

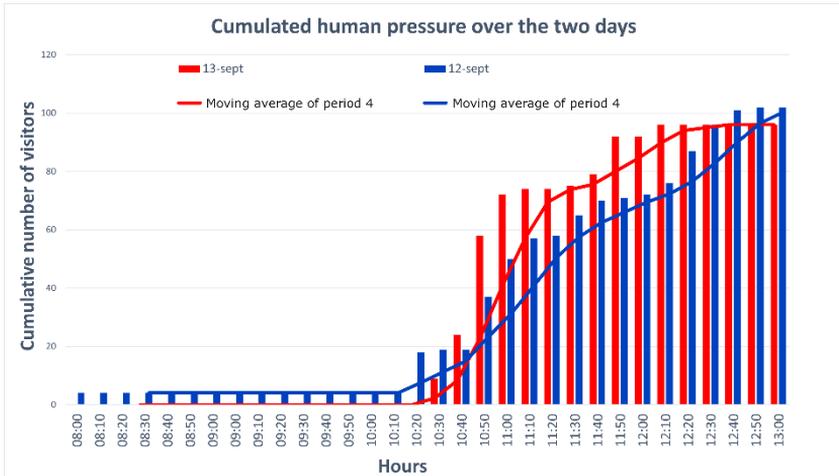

**Figure 3:** *Comparison of the number of cumulated visits over the two mornings (12 September and 13 September 2016). The trends show an equivalence of both series.*

The network of sound sensors (two TASCAM DR40 recorders, 96 kHz SR, 24 bits) were installed on different sections of the trails in order to obtain optimal acoustic recordings (2 recording posts upstream from 9:00 to 20:00; one on the edge of the trail and the other in the undergrowth).

In addition to the answers to the questions, the researchers gathered all the visitor traffic data (time stamping, number of people) to correlate with the acoustic data.

The visitor numbers are similar on both days and both path: the trend of the curves cumulating the number of people passing through in the morning is presented in Fig. 3 below (Pearson correlation: 0.97) and shows comparable human pressure over these two days.

## Results and discussion

We calculated the centre frequency of each hour of recording (Fig. 4). On 12 September, the 'non-binding' information day, the centre frequencies curve is different to that of the following day, the 'binding' information day. The centre frequencies on 12 September fluctuate between 7 kHz +/- 3 kHz (green line shows the tendency). No tourist condition after 17:00 the 16th, give the normal centroid frequency of the soundscape centroid near 8 kHz.

Soundscape at sunrise and sunset is full of fauna songs, with lot of chirps, wide band bird songs, this may explain that the centroid frequency is higher until 9:00 am and after 4:00 pm, up to 16 kHz. Not that the first day, only the morning includes such centroid frequencies. The calm day, the 13th, has such frequencies at the morning and at the evening, the fauna may have not been such impacted than a noisy day (as the 12th). However, on 13 September they are stable around 7 kHz (orange line). This marked difference in the soundscape cannot be attributed to the weather conditions which were almost the same between the 12 and 13 September (+/- 1 degree, same cloud cover) but may be an effect of human voices.

Moreover, we observe that visitors begin to arrive on the site from 10:00, with a high number returning at the end of the afternoon around 16:00. These times are correlated with the falls in centre frequencies on 12 September (non-binding questionnaire), while the same pattern in terms of numbers and time of visitors having completed the 'binding questionnaire' on 13 September does not disturb this acoustic landscape.

The comparison of the two protocols put in place of non-binding eco-awareness versus binding eco-awareness therefore leads us to observe a variation in the sound environment from one protocol to another. In non-binding condition the 12 September tourists are speaking loudly. Then the central frequency of the soundscape is converging to speech frequencies, that are near 4 kHz, near 10:00 and 15:00 when maximum of tourist are present. In binding condition, the 13th sept. tourists are less impacting the natural soundscape, its centroid frequency is stable, as the one in the non-binding condition without tourist (after 17:00 the 12 September).

This variation tends to validate our central hypothesis of a 'change of attitudes and behaviour in term of noise generation through the adoption of the binding communication protocol combining awareness and commitment approaches'.

Furthermore, having also observed that visitors having been made eco-aware through the binding communication procedure made less impact in terms of noise, we conclude that this procedure improves eco-awareness echoing the results of other research studies (Grandjean and Guéguen, 2011; Parant *et al.*, 2017; Terrier and Marfaing, 2015).

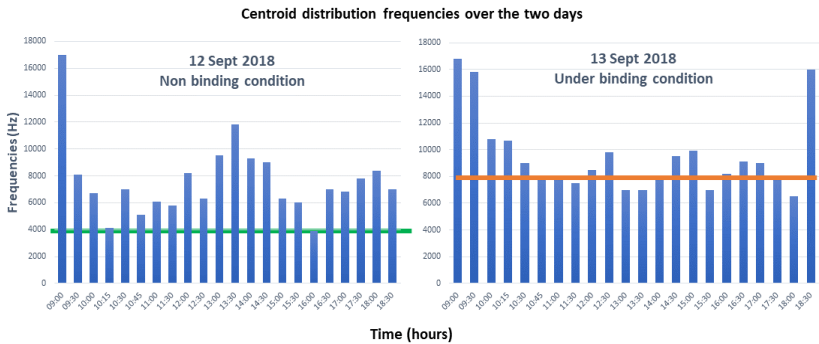

**Figure 4:** *Distribution of centroid frequencies over the two days: the centre frequency of each hour of recording*

Tab. I shows some additional data from the analysis of the questionnaires. The samples of visitors over the two days are approximately the same size and present the same sociodemographic profiles (age, gender, etc.). They also present the characteristics of the chosen period, low season and outside the school holidays ('older' couples without children).

**Table I:** *Some additional results from the analysis of the questionnaires. PACA region: Provence, Alps and French Riviera.*

|  | **Non-binding questionnaire (13/09)** | **Binding questionnaire (12/09)** |
|---|---|---|
| **Sample** | 103 | 103 |
| **Gender** | 52 men, 47 women 4 unknowns | 47 men, 52 women, 4 unknowns |
| **Age / Socio-economic status** | Aged 45-65 / High | Aged 45-65 / High |
| **Origin** | Mainly PACA region, Lyon, Paris, Switzerland, Belgium, Germany | Mainly PACA region, Lyon, Paris, Switzerland, Belgium, Germany |
| **Act of filling in the questionnaire** | 88% completed in full but often not very neatly | 58% completed in full and neatly |
| **Number committed** | Not relevant | 92/103 (box ticked) of whom 63 completed it fully |
| **Question regarding the awareness of their noise disturbance** | Average scores (1 to 10) = 5.78 | Average scores (1 to 10) = 5.81 |

Awareness of the human impact on a protected site is the same for both groups (scores of 6 on a scale of 1 to 10).

We were also able to observe the effects of commitment on visitors through an unexpected result: visitors from the 'binding communication' group took greater care when filling in the questionnaire than the others. We think that this behaviour of the visitors of the Port-Cros Island are more suitable to receive messages or rules and furthermore follow the environmental one joining results in previous study in the National Park (Le Berre *et al.*, 2013).

Our experimental protocol has a positive effect on visitors' behaviour in terms of their noise emissions. Simply ticking a box at the end of the attitude questionnaire for the 'binding awareness' group was enough to make these visitors adopt quiet behaviour. This result suggests that 'foot in the door' techniques (Joule and Beauvois, 2002) are efficient in the area of environmentally friendly behaviour. While this type of preparatory act encourages visitors to a natural park to adopt eco-friendly actions, it is also an essential preliminary step towards the spreading of a "culture of responsibility" (Bernard, 2006).

**Conclusion**

The preliminary results obtained from this study enable us to contribute to the relevance, effectiveness and feasibility of a binding communication protocol in the area of eco-citizenship.

The results and the conclusions (Duvernay *et al.*, 2018) of this project are encouraging, in the sense that they enable us to validate an innovative approach with respect to synergy and coordination between researchers belonging to two distinct scientific fields: Social Sciences and exact and formal sciences.

With regard to the limitations of our work, we note the absence of a control group in the conditions initially intended, which would have enabled us to reinforce our conclusions. Despite this methodological limit, we consider, given the collaboration conditions and the results obtained, that this experiment contributes to the epistemological, theoretical, methodological and empirical debate relating to 'methodological pluralism', which gives equal consideration to the competencies of researchers, economic stakeholders and citizens.

**Perspectives**

The main joint research protocol offers a way of setting up comparative performance communication tests and provides a starting point for more extensive multidisciplinary work which may arouse the interest of the scientific community and civil society, with regard to contemporary eco-citizen issues. The whole protocol aims to devise a way of raising visitors' awareness by separating different methods of communication and assessing their effectiveness thanks to bioacoustic metrics. Future work will consider a variety of awareness-raising methods implemented across several tourist profiles (according to different social criteria: gender, education, origin, etc.) in order to establish the best communication methods for educating people and minimizing human impact on ecosystems wherever possible.

In September 2018, we repeated this experiment for the "Captiles 2" project in order to consolidate our initial results and to extend our analysis protocols:

- bioacoustic detection of anthropogenic disturbance which, instead of measuring the impact on the acoustic landscape,

measures the noise generated by visitors. This is a new bioacoustics challenge

- a set of communication protocols to distinguish between different awareness-raising methods, supplemented by questionnaires to determine social characteristics
- protocols are combined with an assessment of the effectiveness of awareness-raising actions through the measurement of the anthropogenic impact.

The purpose of the bioacoustic analysis is to determine any potential discrepancies in noise pollution that can be observed between the two days. Any differences will consequently serve as a measure of the effectiveness of the communication actions. This will be modulated according to the number of 'customers', and their characteristics (gender, social-economic status, interests).

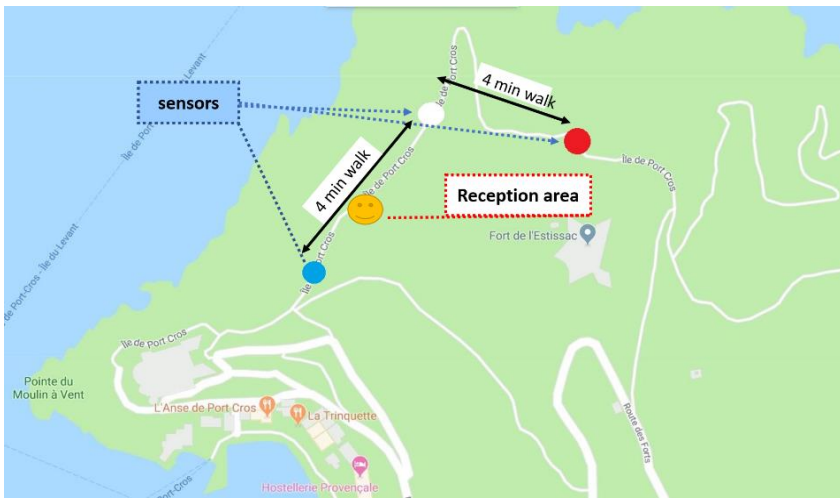

**Figure 5:** *Position of the sensors and path 1 of the study area, the blue and white sensors are in virtually the same position as in the 2016 experiment*

This year two paths (Fig. 2) were used during our tests to ensure that the results were not dependent on a particular area.

We used three stereo sound recorders, 16-bits, 96 kHz for each path. Each recorder was placed about a four-minute walk from each other as seen in Fig. 5 and 6.

The experiment was carried out over 4 days: from Tuesday 18 to Friday 21 September 2018, each day focusing primarily on a specific data-gathering process.

- day 1: to capture the reference 'mood', we recorded visitor traffic

- day 2: we displayed some posters asking visitors to remain silent and recorded visitor traffic
- day 3: we gave visitors eco-awareness information and recorded visitor traffic
- day 4: we gave visitors eco-awareness information and urged them to respect the silence using binding communication methods. We devised a way of carrying out group awareness and commitment actions (used only on the second path).

The on-site team comprised five researchers, as well as 3 trainees to oversee the protocols and administer the questionnaires. We are now processing the data gathered (more than 1000 questionnaires and 100 Go of sound recording) and expect the results soon. The combined implementation of the two protocols enables an interdisciplinary approach to the mutual qualification of the data to be established. The results can potentially lead to the development of instruments measuring in real time the load of the effective anthropic pressure on fauna in an ecosystem. This work, particularly in view of its innovative approach combining bioacoustics and information and communication sciences, opens up an unprecedented operational path for the implementation of eco-awareness actions.

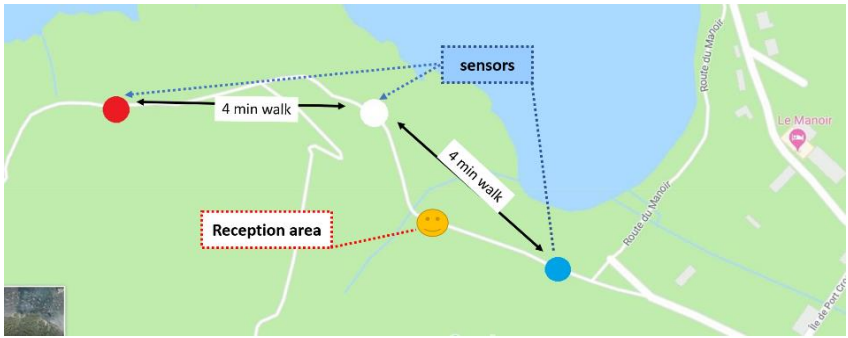

**Figure 6:** *Position of the sensors and the reception area on the second path*

**Acknowledgments:** We wish to thank The Université de Toulon and the Métropole Toulon-Provence-Méditerranée for the funding, The Port-Cros National Park for their logistics support, The *Information Numérique Prévention Santé* (INPS) and *Mer Environnement et Développement Durable* (MEDD) research centers of the Université de Toulon for editing and supervising this research project. Thanks also to the reviewers and to Prof. Charles-François Boudouresque for their valuable comments.